\documentclass{cs19proc}
\usepackage{amsmath, color, hyperref}
\usepackage{titlesec}

\newcommand{\vB}{\boldsymbol{B}}
\newcommand{\bxi}{\boldsymbol{\xi}}
\newcommand{\bpsi}{\ensuremath{\pmb{\psi}}}
\newcommand{\grad}{\boldsymbol{\nabla}}
\newcommand{\bcdot}{\!\boldsymbol{\cdot}\!}
\newcommand{\dvg}{\boldsymbol{\nabla}\!\bcdot}
\newcommand{\curl}{\boldsymbol{\nabla}\!\!\boldsymbol{\times}\!\!}
\newcommand{\cnabla}{\!\boldsymbol{\cdot}\!\!\boldsymbol{\nabla}}
\newcommand{\cross}{\!\!\boldsymbol{\times}\!\!}
\newcommand{\rht}{\hat{\mathbf{r}}}

\editors{G.~A. Feiden}
\publisher{Zenodo}
\conference{The 19th Cambridge Workshop on Cool Stars, Stellar Systems, and the Sun}
\conferencedate{2016}

\title{On the Stability of a General Magnetic Field Topology in Stellar Radiative Zones}
\author{Kyle Augustson$^1$, St\'{e}phane Mathis$^{1,2}$, Antoine Strugarek$^{3,1}$}

\affiliation{$^1$Laboratoire AIM Paris-Saclay, CEA/DRF -- CNRS -- Universit\'{e} Paris Diderot, IRFU/SAp Centre de Saclay, F-91191 Gif-sur-Yvette Cedex, France \\
  $^2$LESIA, Observatoire de Paris, PSL Research University, CNRS, Sorbonne Universit\'{e}s, UPMC Univ. Paris 06, Univ. Paris Diderot,
  Sorbonne Paris Cit\'{e}, 5 place Jules Janssen, 92195 Meudon, France \\
  $^{3}$D\'{e}partement de physique, Universit\'{e} de Montr\'{e}al, C.P. 6128 Succ. Centre-Ville, Montr\'{e}al, QC H3C-3J7, Canada}

\shorttitle{Generalized Tayler Instability}
\shortauthors{Augustson et al.}

\abs{This paper provides a brief overview of the formation of stellar fossil magnetic fields and what potential
    instabilities may occur given certain configurations of the magnetic field.  In particular, a purely magnetic
    instability can occur for poloidal, toroidal, and mixed poloidal-toroidal axisymmetric magnetic field configurations
    as originally studied in \citet{tayler73}, \citet{markey73}, and \citet{tayler80}. However, most of the magnetic
    field configurations observed at the surface of massive stars are non-axisymmetric. Thus, extending earlier studies
    of the axisymmetric Tayler instability in spherical geometry \citep{goossens80}, we introduce a formulation for the
    global change in the potential energy contained in a convectively-stable region given an arbitrary Lagrangian
    perturbation, which permits the inclusion of both axisymmetric and non-axisymmetric magnetic fields. With this tool
    in hand, a path is shown by which more general stability criterion can be established.}

\begin{document}

\maketitle

\section{Motivation}

The radiative core of main-sequence low-mass stars and the radiative envelope of main-sequence massive stars likely host
a fossil magnetic field \citep{neiner15,braithwaite15}. This field is a remnant of the field built during the star's
birth and subsequently reinforced during convective phases of its evolution toward the main-sequence
\citep{alecian13}. In particular, massive stars with an observed magnetic field typically possess a non-axisymmetric
oblique magnetic dipole or a similarly simple magnetic field geometry \citep{moss90,walder12,wade16}. If a
comparison is drawn between the stably-stratified regions of massive and low-mass stars \citep{strugarek11}, given their
hydrodynamic similarity, such non-axisymmetric magnetic fields may also exist within these regions for low-mass
stars. Fossil magnetic fields have also been proposed as an important source of angular momentum transport and mixing
across the Hertzsprung-Russell diagram \citep[e.g.,][]{gough98,heger05,mathis05}. So, constraining the stability of a
large class of magnetic fields within the convectively-stable, radiative regions is important for characterizing their
influence on the transport of angular momentum over evolutionary timescales, understanding their topology that is
observed at the surfaces of intermediate and high mass stars, and their consequences for the local stellar environment
\citep[e.g.,][]{petit12}.

\begin{figure*}[!t]
	\centering
	\includegraphics[width=0.85\linewidth]{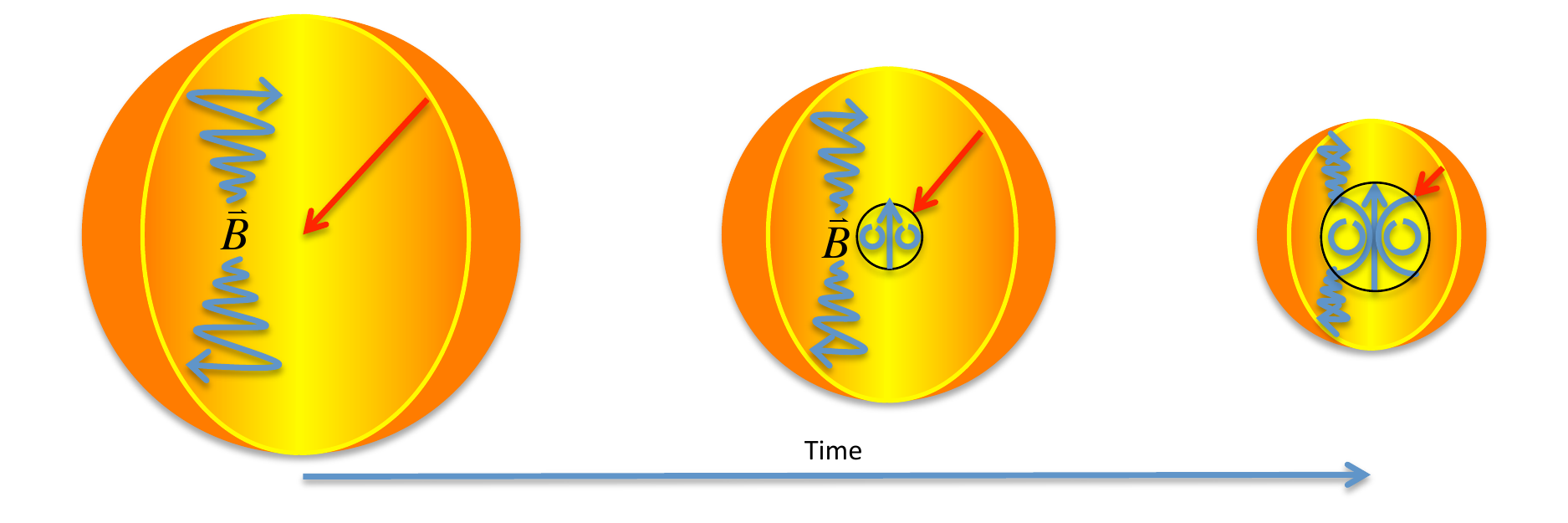}
	\caption{Transition from an initially fully convective, pre-main-sequence star to a main-sequence star with a
          stable radiative interior. The fossil field results from the magnetic field established by the convective
          dynamo, but once convection has halted, it relaxes into a stable configuration during the evolution of
          the stable region.  The stellar magnetic field is a superposition of dynamo-generated and fossil fields. The
          red arrow denotes the contraction of the convective envelope, the convectively-stable core is the yellow
          region encirled by a black line.}
	\label{fig:relax}
\end{figure*}

As an example of how such fossil fields can form, the process of freezing out the magnetic field as the star evolves
along the pre-main-sequence is depicted in Figure \ref{fig:relax}, where gravitational contraction decreases the radius
of the star.  As the star slowly collapses, the gradual increase of the density and temperature deep within the star
tends to lower the opacity, which eventually leads transition from convective heat transport to diffusive heat
conduction at the edge of the core. During these phases, rotationally-constrained convective motions will generate the
magnetic field. In contrast, once convection has halted in the stably-stratified layers, the field will undergo a slow
Ohmic decay if the field has a stable configuration or a fast Alfv\'{e}nic decay if it is unstable. One way to
distinguish which of these decay paths the magnetic field will take is to assess its stability to small (linear)
displacements of fluid elements.  If the growth rate of those small perturbations is real and positive, the
magnetic field undergoes the Tayler instability.

\section{The Tayler Instability}

\begin{figure}[!t]
	\centering
	\includegraphics[width=\linewidth]{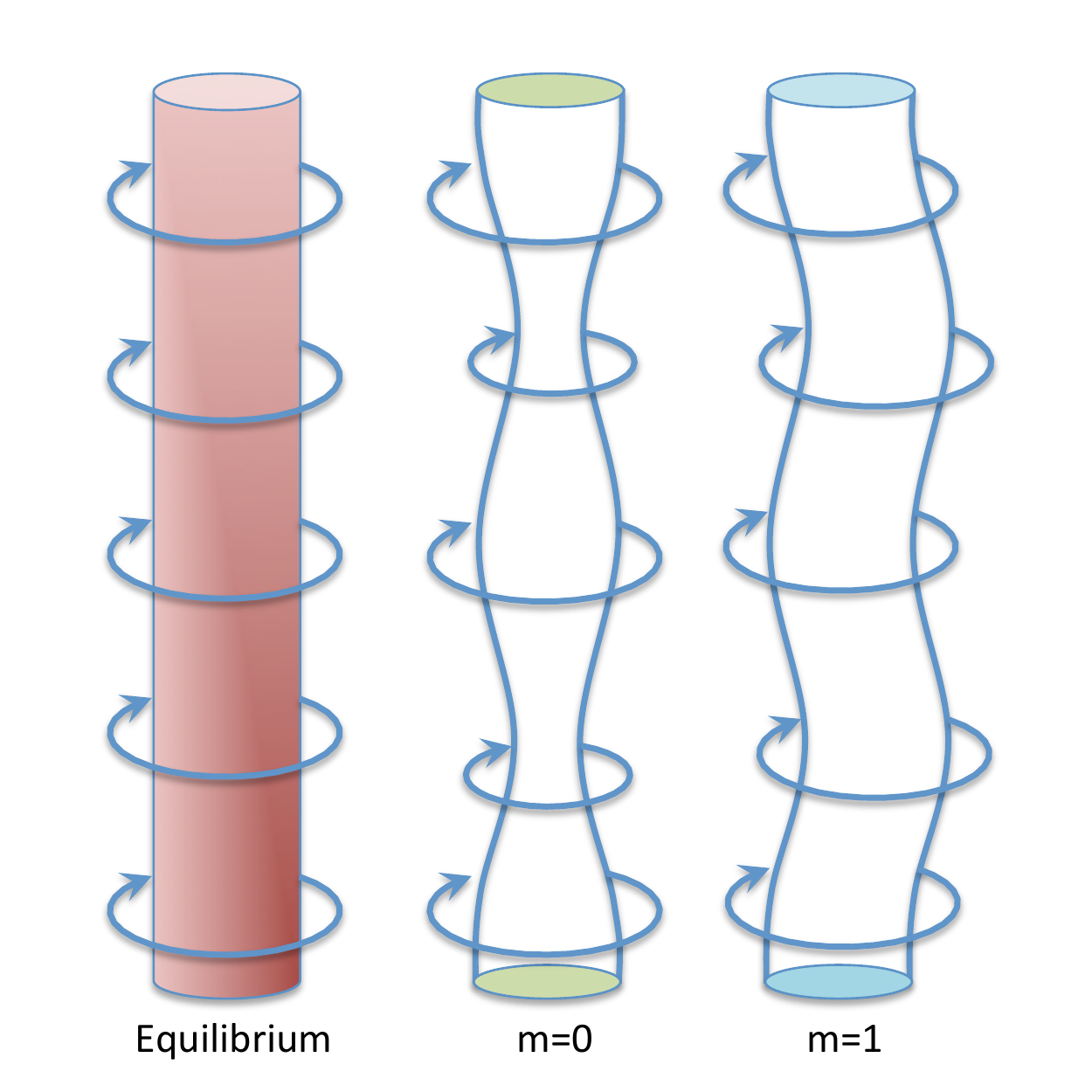}
	\caption{Tayler instabilities in a cylindrical current channel. From left to right: shows an equilibrium configuration
          for an azimuthal magnetic field, a varicose (m=0) instability, and a kink type (m=1) instability.  Field lines
          are marked with arrows.}
	\label{fig:tayler}
\end{figure}

The stability of axisymmetric magnetic field configurations within a quiescent, stably-stratified medium have been
understood for quite some time, with \citet{tayler73} addressing toroidal field configurations and \citet{markey73}
poloidal configurations. The typical local instabilities arising in those systems are shown in Figure \ref{fig:tayler},
where there are three situations shown.  The equilibrium situation occurs when the magnetic field has no associated
current (e.g., if it is potential), or if the Tayler stability criterion are met.  Yet for sufficiently large currents
or sufficiently strong Lorentz forces, two other instabilities can be excited: the axisymmetric $m=0$ varicose
instability, or the $m=1$ kink instability.  The latter of which grows most rapidly when excited. Furthermore, such
analyses indicate that only certain mixed (poloidal and toroidal) configurations of axisymmetric magnetic fields are
stable within the radiative regions of stars \citep{tayler80, braithwaite09}. The presence of rotation modifies the
stability characteristics of these axisymmetric systems in that it tends to further stabilize them through the Coriolis
force \citep{pitts85}. The precise form of the equilibrium states of the mixed-morphology magnetic fields has been
considered extensively in both non-rotating and rotating systems
\citep{prendergast56,braithwaite06,braithwaite08,duez10a,duez10b,duez11,braithwaite13,emeriau15}.

\section{Generalizing the Tayler Instability}

The class of local stability analysis established by \citet{tayler73} can be generalized to global-scale geometries as
can be found in \citet{goossens80}.  However, axisymmetric magnetic fields are not the final story in the study of the
Tayler instability. Rather, the analysis can be extended to configurations with both non-axisymmetric magnetic fields
and differential rotation as shall be shown in an upcoming paper \citep{augustson16}. The resulting stability criteria
are assessed here. Such criteria help to restrict the number of magnetic field configurations that are possible within
the stable regions of low-mass stars, thereby limiting the routes of angular momentum transport in the radiative
interior and means of interaction with the dynamo-generated magnetic fields established in their overlying convective
layers.\\

The linearized equation of motion under the Cowling approximation \citep{cowling41} for a fluid element in a general,
but non-rotating, coordinate system is

\vspace{-0.25truein}
\begin{center}
  \begin{align}
    \rho\frac{\partial^2 \bxi}{\partial t^2} &= \frac{1}{4\pi}\left[\left(\curl{\delta\vB}\right)\cross\vB+\left(\curl{\vB}\right)\cross\delta\vB\right]-\delta\rho\grad{\Phi}-\grad{\delta P}, \label{eqn:leom}
  \end{align}
\end{center}

\noindent where $\bxi$ is the displacement, $\vB$ the magnetic field, $\rho$ the density, $P$ the pressure, $\Phi$ the
gravitational potential. The Eulerian perturbations $\delta$ of those quantities follow directly from the continuity,
pressure, and induction equations as

\vspace{-0.25truein}
\begin{center}
  \begin{align}
    \delta\rho &= -\dvg{\left(\rho\bxi\right)},\\
    \delta P &= -\bxi\cnabla P - \gamma P \dvg{\bxi},\\
    \delta \vB &= \curl{\left(\bxi\cross\vB\right)}.
  \end{align}
\end{center}

\noindent Therefore, one has that

\vspace{-0.25truein}
\begin{center}
  \begin{align}
    \rho\frac{\partial^2\bxi}{\partial t^2} &= \mathcal{F}[\bxi; \rho, P, \Phi, \vB] \nonumber \\
    &= \dvg{\left(\rho\bxi\right)}\grad{\Phi}+\grad{\left[\bxi\cnabla P + \gamma P \dvg{\bxi}\right]} \label{eqn:full} \\
    &+\left[\left(\curl{\curl{\left(\bxi\cross\vB\right)}}\right)\cross\vB+\left(\curl{\vB}\right)\cross\left(\curl{\left(\bxi\cross\vB\right)}\right)\right], \nonumber
  \end{align}
\end{center}

\noindent where $\gamma$ is the ratio of specific heats.

As was shown in \citet{bernstein58} and to decide upon the stability of this system, one can consider simple
solutions of the form $\bxi = Re\left[\bpsi(\mathbf{x}) \exp{\left(i\omega t\right)}\right]$, for which the
equation of motion yields $-\omega^2 \rho\bpsi = \mathcal{F}\left[\bpsi\right]$.  This is not general since one has not
yet proven that these basis functions form a complete set on the Hilbert space for the Eulerian system. Yet it can be
shown that the vector function $\mathcal{F}$ is self-adjoint, a proof of which will be reserved for the upcoming paper
\citep{augustson16}. With a properly defined inner product for the solutions $\bxi$, one can see that the dispersion
relationship for a general displacement in an arbitrary coordinate system is given by

\vspace{-0.25truein}
\begin{center}
  \begin{align}
    \omega^2 &= -\frac{\langle\bpsi,\mathcal{F}\left[\bpsi\right]\rangle}{\langle\rho\bpsi,\bpsi\rangle}
             = -\int \bpsi^{*}\bcdot\mathcal{F} dV \left[\int \rho \bpsi^{*}\bcdot\bpsi dV\right]^{-1},  \nonumber \\
             &= 2\Delta W \left[\int \rho \bpsi^{*}\bcdot\bpsi dV\right]^{-1},
  \end{align}
\end{center}

\noindent where the integral is taken over the region of interest, which for stars are their convectively-stable zones.

The displacement is unstable if the change in the potential energy of the system ($\Delta W$) is negative. In general,
one finds that this energy can be split into three parts as $\Delta W = \Delta W_L + \Delta W_B + \Delta W_P$, with
$\Delta W_L$ being the work due to Lorentz forces, $\Delta W_B$ being the work due to buoyancy, and with $\Delta W_P$
being the pressure work.  To find general classes of magnetic fields that are stable in a given radiative region, one
needs to be able to find an expression for the work that can be minimized.  This is possible within the context of
separable coordinate systems.  For this work, the spherical coordinate system is used.  Therefore, for compactness and
expedience, $\bpsi$, $\rho$, $P$, $\Phi$, and $\vB$ are projected onto the spherical spin vector harmonics (SVH). The
SVH are a complete orthonormal set of vector functions that are formed from specific combinations of the spherical
harmonics and their derivatives \citep{varshalovich88}.  In particular, they correspond to the joint eigenstates of the
angular momentum and spin-1 operators.  Relative to the RST basis \citep{rieutord87}, vector operations such as the dot
and cross products are simpler to perform on SVH-projected vector-valued functions. An explicit representation of the
SVH in terms of scalar spherical harmonics is as follows:

\vspace{-0.25truein}
\begin{center}
  \begin{align}
    \mathbf{Y}_{\ell,1}^m &= \frac{1}{\sqrt{\left(\ell+1\right)\left(2\ell+1\right)}}\left[-\left(\ell+1\right)\rht + r\grad{}\right] Y_\ell^m, \label{eqn:svhp1}\\
    \mathbf{Y}_{\ell,0}^m  &= \frac{-i r}{\sqrt{\ell\left(\ell+1\right)}}\rht\cross\grad{Y_\ell^m}, \label{eqn:svhp0}\\
    \mathbf{Y}_{\ell,-1}^m &= \frac{1}{\sqrt{\ell\left(2\ell+1\right)}}\left[\ell\rht + r\grad{}\right] Y_\ell^m \label{eqn:svhm1},
  \end{align}
\end{center}

\noindent where the second lower index $\nu$ on the $\mathbf{Y}_{\ell,\nu}^m$ indicates the corresponding spin-1 basis
vector.  Indeed, when the work integrands are expanded on the SVH basis, one can show that 

\vspace{-0.25truein}
\begin{center}
  \begin{align}
    &\Delta W_L = \!\!\int_{r_b}^{r_t} \!\! dr r^2\sum_{i=1}^4\sum_{\substack{\ell,m,\nu,\mu,\lambda \\ \ell_i, m_i,
    \nu_i}} w^{\mu,\lambda}_{\substack{\ell,m,\nu \\ \ell_i, m_i, \nu_i}},
  \end{align}
\end{center}

\noindent with,

\vspace{-0.25truein}
\begin{center}
  \begin{align}
    &\sum_{i=1}^4 w^{\mu,\lambda}_{\substack{\ell,m,\nu \\ \ell_i, m_i, \nu_i}}\!\!\!=\nonumber \\ 
    &{_{\lambda,\mu}L_{\substack{\ell_1,\ell_2\\ \nu_1, \nu_2}}^{m_1, m_2}}\!\!\left(\psi_{\ell_3,\nu_3}^{m_3}B_{\ell_4,\nu_4}^{m_4}\right)\psi_{\ell,\nu}^{*m} 
    \mathcal{J}_{\substack{\ell_3, m_3, \nu_3 \\ \ell_4, m_4, \nu_4}}^{\ell_2, m_2, \lambda}
    \mathcal{J}_{\substack{\ell_1, m_1, \nu_1 \\ \ell_2, m_2, \nu_2}}^{\ell, m, \nu},
  \end{align}
\end{center}

\noindent where each $\ell$ ranges from zero to infinity, each $m$ ranges between $-\ell$ and $\ell$, and where $\nu$,
$\mu$, and $\lambda$ range between $-1$ and $1$. The integral is taken between radii $r_b$ and $r_t$, which demark the
bottom and top boundaries of the radiatively stable region. The $\mathcal{J}$ coefficients arise from the projection of
the cross products of the SVH basis vectors in Equation \ref{eqn:full} back onto the basis. The $L$ symbol is a function
of radius that arises from the Lorentz force, and thus it is a second-order differential operator involving the radial
functions of the displacement and the magnetic field as

\vspace{-0.25truein}
\begin{center}
  \begin{align}
    &_{\lambda,\mu}L_{\substack{\ell_1,\ell_2 \\ \nu_1, \nu_2}}^{m_1, m_2}\!\left(\psi_{\ell_3,\nu_3}^{m_3} B_{\ell_4,\nu_4}^{m_4}\right)\! = \nonumber \\
    &\!-\frac{B_{\ell_1, \nu_1}^{m_1}}{4\pi}\left[E_{\nu_2,\lambda}^{\ell_2, m_2}\frac{\partial^2}{\partial r^2} 
      + \frac{F_{\nu_2,\lambda}^{\ell_2, m_2}}{r}\frac{\partial}{\partial r}
      + \frac{G_{\nu_2,\lambda}^{\ell_2, m_2}}{r^2}\right]\psi_{\ell_3,\nu_3}^{m_3} B_{\ell_4,\nu_4}^{m_4}\mathcal{I}_{\lambda,\mu} \nonumber\\
    &+\frac{1}{4\pi}\left[D_{\nu_1,\lambda}^{\ell_1,m_1} \frac{\partial B_{\ell_1,\lambda}^{m_1}}{\partial r} 
      + C_{\nu_1,\lambda}^{\ell_1,m_1}\frac{B_{\ell_1,\lambda}^{m_1}}{r}\right]\times \nonumber \\
    &\qquad\qquad\qquad\left[D_{\nu_2,\mu}^{\ell_2,m_2}\frac{\partial}{\partial r}+\frac{C_{\nu_2,\mu}^{\ell_2,m_2}}{r}\right]\psi_{\ell_3,\nu_3}^{m_3}B_{\ell_4,\nu_4}^{m_4},
\end{align}
\end{center}

\noindent where the coefficient matrices $C$, $D$, $E$, $F$, and $G$ describe the projection of the curl and double curl
operators onto the spin vector harmonic basis, and $\mathcal{I}$ is the unit tensor.  Assuming that the star is
spherically symmetric, namely that the gradient of the gravitational potential is only in the radial direction, then one
has that $g=-\partial_r{\Phi}$.  So, tackling the buoyancy work integral, it can be seen that

\vspace{-0.25truein}
\begin{center}
  \begin{align}
    &\Delta W_B = \!\!\!\!\sum_{\substack{\ell, m, \ell_1 \\ \ell_2, m_2,\nu_2}} \!\!\!\int_{r_b}^{r_t} \!\! dr \frac{\left(-1\right)^{m_2} g r^2}{2\ell+1}
    \left(\sqrt{\ell+1}\psi_{\ell,1}^{*m}- \sqrt{\ell}\psi_{\ell,-1}^{*m}\right) \nonumber \\
    &\left[\sqrt{\ell+1}\left(\frac{\partial}{\partial r}+\frac{\ell+2}{r}\right)\mathcal{K}_{\ell_1;\substack{\ell,1\\ \ell_2,\nu_2}}^{-m, m_2}\right.\nonumber\\
    &\qquad\qquad\left.-\sqrt{\ell}\left(\frac{\partial}{\partial r}-\frac{\ell-1}{r}\right)\mathcal{K}_{\ell_1;\substack{\ell,-1\\ \ell_2,\nu_2}}^{-m, m_2}\right]
    \rho_{\ell_1}^{m_2-m} \psi_{\ell_2,\nu_2}^{m_2}.
  \end{align}
\end{center}

\noindent Similarly, the pressure work integral can be identified as

\vspace{-0.25truein}
\begin{center}
  \begin{align}
    &\Delta W_P = 
      \!\!\int_{r_b}^{r_t} \!\! dr r^2\!\!\!\sum_{\substack{\ell,m,\ell_1 \\ \ell_2,m_2}}\frac{1}{\sqrt{\left(2\ell+1\right)\left(2\ell_2+1\right)}}\nonumber\\
    &\left[\sqrt{\ell+1}\left(\frac{\partial}{\partial r} +\frac{\ell+2}{r}\right) \psi_{\ell,1}^{*m+m_2}\right. \nonumber \\
    &\qquad\qquad+\!\left.\sqrt{\ell}\left(-\frac{\partial}{\partial r} +\frac{\ell-1}{r}\right)\psi_{\ell,-1}^{*m+m_2} \right]\nonumber\\
    &\left\{\sum_{\nu_1}\Bigg(\psi_{\ell_1,\nu_1}^{m}\left[\!\sqrt{\ell_2+1}\left(\frac{\partial }{\partial r} 
      - \left.\frac{\ell_2}{r}\right) \mathcal{K}_{\ell;\substack{\ell_1,\nu_1\\ \ell_2,1}}^{m, m_2} \right. \right. \right.\nonumber\\
    &\left. - \sqrt{\ell_2}\left(\frac{\partial }{\partial r} + \frac{\ell_2+1}{r}\right) 
      \mathcal{K}_{\ell;\substack{\ell_1,\nu_1\\\ell_2,-1}}^{m, m_2}\right] P_{\ell_2}^{m_2}\Bigg)\nonumber\\
    &+\left(-1\right)^{m+m_2}\gamma \mathcal{H}_{\substack{\ell,\ell_1,\ell_2\\ m, m_2}}P_{\ell_1}^m
    \left[\sqrt{\ell_2+1}\left(\frac{\partial}{\partial r} +\frac{\ell_2+2}{r}\right) \psi_{\ell_2,1}^{m_2}\right. \nonumber \\
    &\left.+\sqrt{\ell_2}\left(-\frac{\partial}{\partial r}+\frac{\ell_2-1}{r}\right)\psi_{\ell_2,-1}^{m_2} \right]\Bigg\}.
  \end{align}
\end{center}

\noindent Here, $\mathcal{H}$ and $\mathcal{K}$ are coefficients related to the 3-j and 6-j symbols that arise from
integrals over products of SVH that are then either projected onto the scalar spherical harmonics, which, along with
$\mathcal{J}$, are closely related to those defined in \citet{varshalovich88} and \citet{strugarek13}.

\section{Conclusions}

With the expanded form $\Delta W$ in hand, one can then find conditions under which the system with a chosen general
magnetic field is linearly stable or unstable to an arbitrary displacement by integrating the terms with radial
derivatives of $\bpsi$ by parts and then explicitly minimizing the radial integrals with respect to $\bpsi$.  This will
be demonstrated more completely in an upcoming paper \citep{augustson16}.  As applied to stellar radiative zones, this
will permit the determination of the stability of certain classes of magnetic fields that have a broad range of
non-axisymmetric components.  In particular, it may be possible to assess why the magnetic field configuration where the
magnetic axis of symmetry is oblique to the rotation axis of the star is the most commonly observed.

\section*{Acknowledgments} {K.~C. Augustson and St\'{e}phane Mathis acknowledge support from the ERC SPIRE 647383
  grant. A.~Strugarek acknowledges support from the Canadian Institute of Theoretical Astrophysics (National Fellow) and
  from the Canadian Natural Sciences and Engineering Research Council.}

\bibliographystyle{cs19proc}
\bibliography{tayler}

\end{document}